\documentclass[a4paper,11pt]{article}
\usepackage{jcappub}
\usepackage{amsmath,amssymb}
\usepackage{graphicx}
\usepackage{hyperref}
\usepackage{booktabs}
\usepackage[margin=2cm]{geometry}
\usepackage{orcidlink}
\usepackage{titlesec}

\title{Vortex-reconnection energy bounds in Bose-Einstein-condensed and superfluid dark matter halos}

\author[a]{K.~Yavuz~Ek\c{s}i \orcidlink{0000-0001-5999-0553}}

\affiliation[a]{Istanbul Technical University, Faculty of Sciences and Letters, Department of Physics, 34469 Istanbul, Turkiye}

\emailAdd{eksi@itu.edu.tr}

\abstract{
Bose-Einstein-condensed (BEC) and superfluid dark-matter (SFDM) halos can contain coherent, wave-supported cores whose angular momentum is carried by quantized vortices. When vortices form a tangle, reconnections convert part of the vortex kinetic energy into dark phonons, density waves, Kelvin waves, and vortex loops, providing a microscopic channel by which vortex structure can affect halo-core evolution. We estimate the dynamical importance of this channel by combining the local Gross-Pitaevskii (GP) reconnection law with a halo-scale vortex-line density calibrated against Schr\"odinger-Poisson (SP) simulations. Vortex reconnections cannot appreciably restructure a standard SP halo core under the fiducial assumptions: for a $m=10^{-22}\,\mathrm{eV}$, $r_c=1\,\mathrm{kpc}$ soliton and a one-percent energy-transfer efficiency per event, an unforced network transfers at most $1.3\%$ of the core virial energy in $10\,\mathrm{Gyr}$. Reconnection converts ordered vortex energy into sound and smaller-scale excitations; we bound the cumulative transfer using a spectral scale measured in published SP halo simulations. To connect the result with measured systems, we match the standard soliton profile to half-light masses derived with a single mass estimator for eight classical Milky Way dwarf spheroidals. Assigning the measured aperture mass to the soliton gives ceilings of $0.98-4.52\%$ at the Mocz line-density reference and $0.061-3.19\%$ for a modest-rotation normalization. The largest value is a factor of 22 below the core virial-energy scale and limits the fixed-structure velocity change to $2.3\%$. The spectral reference represents fewer than one projected vortex crossing, so these are continuum upper limits; a vortex-free core has no reconnection heating. Reconnections remain worth testing because they are unavoidable when a tangle exists. 
}

\begin{document}
\maketitle
\flushbottom

\section{Introduction}

Cold dark matter describes the large-scale distribution of matter with considerable success \citep{pla+16}. Galactic halos leave more room for alternatives because baryonic feedback, dark-sector microphysics, and individual formation histories all shape their inner structure \citep{bul17,pop17}. Ultralight bosons can alter these scales while preserving the large-scale successes of cold dark matter \citep{hu+00,goo00,sil02,boh07}. For masses $m\sim10^{-22}$--$10^{-20}\,\mathrm{eV}$, their de Broglie wavelength in a galaxy can approach a kiloparsec \citep{boe07,cha11,hui+17}. In the simplest Schr\"odinger-Poisson description, wave pressure then suppresses small-scale structure and supports a solitonic halo core \citep[see Refs.][for reviews]{bul17,hui21,fer21}.

The axion is a natural microscopic candidate. Its cosmological abundance can arise through vacuum realignment \citep{pre+83}. Whether axion dark matter subsequently forms a genuine Bose-Einstein condensate remains debated. Sikivie and Yang \citep{sik09} argued that gravitational rethermalization produces a condensate before galaxy formation. Later work examined this claim critically \citep{dav15,dav13} and challenged it on kinetic-rate grounds \citep{gut+15}. A coherently oscillating classical scalar field can still display much of the same wave phenomenology without establishing thermodynamic condensation.

Superfluid dark matter (SFDM) follows a different route \citep{kho15,ber15,ber16,kho16}. Dark particles form a finite-density condensed phase inside galaxies \citep[see Refs.][for reviews]{kho22,ber+26}, and its phonons couple to baryons. The resulting force can reproduce aspects of modified Newtonian dynamics \citep[MOND; see][for a review]{fam12}, while the cosmological dark matter remains pressureless \citep{kho15,ber15,ber16,kho16}. Bose-Einstein-condensed dark matter (BECDM) and SFDM therefore share a coherent bosonic medium but differ in their dynamics. BECDM relies mainly on wave pressure and self-gravity. SFDM adds an equation of state, a phonon-mediated force, and a baryon coupling \citep{kho+18,her+21,lis+23,fav25}.

Both settings can support quantized vortices. Away from their cores, the flow remains irrotational \citep{lan41}; each singly quantized line carries circulation $\kappa=h/m$ and angular momentum \citep{ons49,fey55,don93,fet01,fet09}. Rotation produces a lattice when vortex formation is energetically favored \citep{abr57,yar+79,fet09}. Tidal torques and mergers can therefore seed vortices in BECDM \citep{sil02,yu02,kai10,zin11,rin12} or SFDM \citep{mau22}, although a slowly rotating non-interacting soliton need not contain a central vortex. Calculations have found disordered vortex structures in non-interacting BECDM halos \citep{moc+17}, possible baryonic kinematic signatures of individual lines \citep{alv+25}, and regular lattices in repulsively self-interacting rotating cores \citep{bra25a,bra25b}. A regular lattice reconnects much less often than a disordered tangle.

Vortex lines reconnect in both classical \citep{kid94,yao22} and quantum fluids \citep{kon21}. Two strands approach, exchange their connectivity, and separate \citep{all+14,bew+08,pao+10,fon+19,min+22,sta+25}. Gross-Pitaevskii dynamics sends part of their ordered kinetic energy into compressible waves and smaller vortex structures \citep{zuc+12,vil+17,gal+19,vil+20}. Repeated events can drain a disordered network even though the closed condensate conserves total energy.

This paper asks how much energy reconnections can transfer within a condensed halo core. We estimate the rate for a pre-existing disordered vortex population and limit the cumulative transfer by the finite energy stored in that population. For the reference $m=10^{-22}\,\mathrm{eV}$, $r_c=1\,\mathrm{kpc}$ soliton, the ten-Gyr ceiling is $1.3\%$ of the virial scale. We then apply the same calculation to eight classical Milky Way dwarf spheroidals by matching an SP soliton to each measured half-light mass. The resulting population ceiling remains below $4.6\%$, or at least a factor of 22 below the core virial-energy scale. This observational mapping is conditional: the published spectral scale normalizes an unknown vortex-line density and does not establish that any one dwarf contains a tangle. A vortex-free core has no reconnection heating, while an ordered lattice requires a separate model for the disturbance that triggers reconnections.

Section~\ref{sec:model} develops the estimate and its domain of validity. Section~\ref{sec:results} presents the parameter scan and the ten-Gyr bound. Section~\ref{sec:discussion} considers the consequences for BECDM and SFDM.

\section{Semi-analytical heating model}
\label{sec:model}

We model a condensed dark-matter core threaded by a disordered vortex population. The estimate combines a local reconnection law with a halo-scale line density and a dimensional rate for a vortex tangle. Schr\"odinger-Poisson simulations \citep{moc+17} provide a spectral length scale, so the line-density normalization remains an explicit parameter.

\subsection{Local reconnection physics}

Vortex reconnections were conjectured by Feynman \citep{fey55} and Schwarz \citep{sch85,sch88}. They have been observed in superfluid helium \citep{bew+08} and atomic condensates \citep{ser+15}, and were already present in early Gross-Pitaevskii simulations \citep{kop93}.

Kelvin's circulation theorem and Helmholtz vortex freezing prohibit reconnections of material vortex lines in an ideal Euler fluid; helicity conservation reflects the same topological constraint \citep{mof69}. Viscosity removes this obstruction in a Navier-Stokes fluid and allows antiparallel vortices to reconnect and form smaller structures \citep{yao22}.

Quantized vortices are phase singularities rather than tubes of continuous vorticity. The Gross-Pitaevskii equation allows their topology to change while conserving the total energy of the closed condensate. Near the reconnection point, part of the ordered kinetic energy becomes sound \citep{zuc+12}. Vortex-filament \citep[VFM, e.g.][]{sch85,wac+14} and Gross-Pitaevskii calculations \citep{kur+11} describe how the sharp cusps relax after the event.

For a singly quantized vortex the circulation is
\begin{equation}
    \kappa = h/m,
\end{equation}
where $m$ is the boson mass. The particle mass enters the local reconnection problem through this circulation quantum. Near a reconnection at $t=t_0$, the distance between the two reconnecting vortex segments follows a square-root law seen in experiments \citep{bew+08} and in both Gross-Pitaevskii and vortex-filament simulations \citep{gal+19},
\begin{equation}
    \delta(t)=A_{\pm} \left(\kappa |t-t_0|\right)^{1/2}.
    \label{eq:stasiak_scaling}
\end{equation}
The coefficient $A_{-}$ describes the approach before the reconnection and $A_+$ describes the separation afterwards. 
The measured asymmetry $A_+>A_-$ means that the vortices separate faster than they approach. Acoustic emission produces this irreversibility at $T=0$ \citep{vil+20}; mutual friction with the normal component modifies it at finite temperature \citep{sta+25}. Equation~\eqref{eq:stasiak_scaling} has been measured in superfluid helium \citep{bew+08,pao+10,sta+25}, reproduced in Gross-Pitaevskii calculations \citep{zuc+12,vil+17,gal+19}, studied in finite-temperature atomic condensates \citep{all+14}, and shown to apply to slender classical vortices \citep{yao22}. We use it as a local scaling law and treat the coefficients $A_\pm$ as dark-matter calibration parameters. 

\subsection{Halo calibration}

\begin{figure}[!htbp]
\centering
\includegraphics[width=0.95\columnwidth]{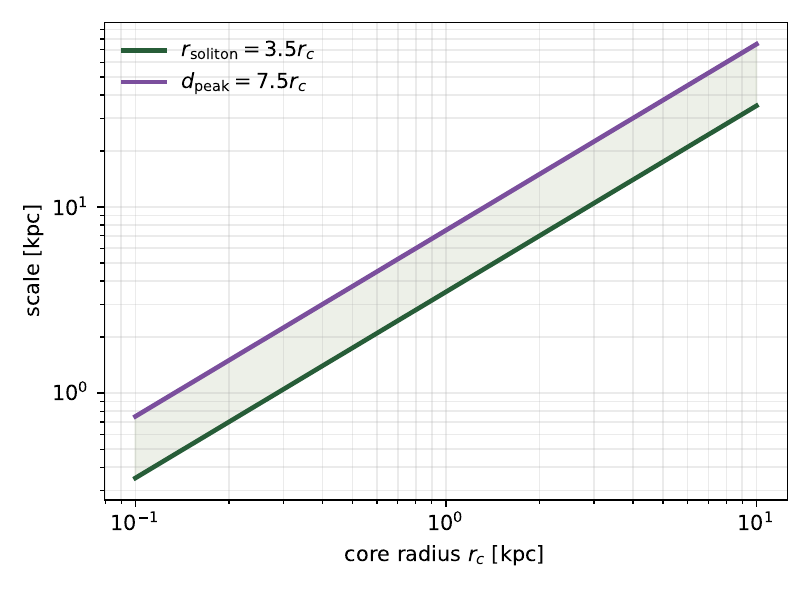}
\caption{The halo scales used in the calculation, taken from the Schr\"odinger-Poisson simulations of Mocz et al. \citep{moc+17}. The soliton size is $r_\mathrm{soliton}=3.5r_{\mathrm{c}}$, and the turbulent spectral peak lies at $d_\mathrm{peak}=7.5r_{\mathrm{c}}$. The spectral peak sets the reference $\mathcal{L}_\mathrm{M}=d_\mathrm{peak}^{-2}$, while the free parameter $\eta_{\rm v}$ sets the physical line density through $\mathcal{L}=\eta_{\rm v}\mathcal{L}_\mathrm{M}$.}
\label{fig:mocz_scaling}
\end{figure}

The local law describes one event, whereas a halo-scale rate also needs the amount of vortex line per unit volume. Mocz et al. \citep{moc+17} report two useful length scales from Schr\"odinger-Poisson simulations of halos with a soliton core and a turbulent outer region. \autoref{fig:mocz_scaling} displays these scales:
\begin{equation}
    r_\mathrm{soliton}\simeq 3.5 r_{\mathrm{c}},
    \qquad
    d_\mathrm{peak}\simeq 7.5 r_{\mathrm{c}},
    \label{eq:mocz_scales}
\end{equation}
where $r_{\mathrm{c}}$ is the usual soliton core-radius parameter, approximately the half-density radius in the standard soliton profile, and $d_\mathrm{peak}$ is the scale at which the turbulent velocity spectrum peaks. We use $d_\mathrm{peak}$ only to define a reference normalization for the unknown vortex-line density. Interpreting this spectral scale as an intervortex scale is the strongest astrophysical assumption in the calculation.

The simulations establish a peak in the turbulent velocity spectrum. A vortex census requires an additional measurement of phase singularities with depleted density cores and countable line length. Wave interference in non-interacting fuzzy dark matter also produces granules, beating modes, and density fluctuations near the local de Broglie scale. The spectral peak may therefore trace granulation instead of an intervortex separation. We use it as a fiducial proxy and retain an explicit normalization for the true line density $\mathcal{L}$, which may lie well below $d_\mathrm{peak}^{-2}$.

The vortex-line density is line length per unit volume. A tangle whose typical separation is $d$ has
\begin{equation}
    \mathcal{L}\sim d^{-2}.
    \label{eq:vortex_line_density}
\end{equation}
We define the Mocz-scale normalization
\begin{equation}
    \mathcal{L}_\mathrm{M}\equiv d_\mathrm{peak}^{-2}=(7.5r_{\mathrm{c}})^{-2},
\end{equation}
and write the unknown line density as
\begin{equation}
    \mathcal{L}=\eta_{\rm v}\mathcal{L}_\mathrm{M},
    \qquad
    b\equiv\mathcal{L}^{-1/2}=\eta_{\rm v}^{-1/2}d_\mathrm{peak}.
    \label{eq:eta_v_definition}
\end{equation}
The case $\eta_{\rm v}=1$ defines the Mocz-normalized reference point. Values below or above unity describe line densities smaller or larger than this proxy, and $\eta_{\rm v}=0$ gives $\Gamma_\mathrm{rec}=Q_\mathrm{rec}=0$. Interpreted as a straight-vortex areal density, $\mathcal{L}_\mathrm{M}$ gives an expected number
\begin{equation}
    N_\perp=\pi r_\mathrm{soliton}^2\mathcal{L}
    \simeq0.68\eta_{\rm v}
    \label{eq:vortex_crossings}
\end{equation}
of crossings through the projected soliton core. The reference spacing $d_\mathrm{peak}=7.5r_\mathrm{c}$ also slightly exceeds the soliton diameter $2r_\mathrm{soliton}=7r_\mathrm{c}$. Thus the reference point contains fewer than one straight core crossing on average. For $\eta_{\rm v}\lesssim1$, the homogeneous-tangle rate below is only a formal ensemble- or time-averaged extrapolation for disturbed cores. A single relaxed core at this normalization generally contains no local tangle to which the Vinen law can be applied. A direct phase-singularity count must ultimately determine $\mathcal{L}$ in the core and turbulent envelope.

\subsection{Rate and heating estimates}

The reconnection rate per unit volume must be built from the circulation quantum $\kappa$ and the line density $\mathcal{L}$ if no other local scale is introduced. Since $\kappa$ has dimensions $\mathrm{length}^2\,\mathrm{time}^{-1}$ and $\mathcal{L}$ has dimensions $\mathrm{length}^{-2}$, the combination with dimensions of events per volume per time is $\kappa\mathcal{L}^{5/2}$. We write
\begin{equation}
    \Gamma_\mathrm{rec}=C_\mathrm{rec}\,\kappa\,\mathcal{L}^{5/2},
    \label{eq:reconnection_rate}
\end{equation}
where $C_\mathrm{rec}$ is a dimensionless coefficient. This scaling is the usual Vinen-type \citep{vin57a,vin57b,vin57c} estimate for the reconnection frequency in a turbulent vortex tangle \citep{bar04,sta+25,bar+23}. The coefficient $C_\mathrm{rec}$ absorbs geometry, polarization of the tangle, and the threshold for counting an event.

The volume of the solitonic core is taken to be
\begin{equation}
    V_\mathrm{core}=\frac{4\pi}{3}r_\mathrm{soliton}^3.
\end{equation}
The total reconnection rate inside this volume is $\Gamma_\mathrm{rec}V_\mathrm{core}$, and hence
\begin{equation}
    t_\mathrm{rec}=\left(\Gamma_\mathrm{rec}V_\mathrm{core}\right)^{-1}.
\end{equation}
The core dynamical time is
\begin{equation}
    t_\mathrm{dyn}\sim (G\rho_\mathrm{c})^{-1/2}.
\end{equation}
This motivates the classifier
\begin{equation}
    \chi \equiv \frac{t_\mathrm{dyn}}{t_\mathrm{rec}}.
    \label{eq:chi}
\end{equation}
If $\chi\ll 1$, reconnections are rare during a core dynamical time. If $\chi\gtrsim 1$, reconnections are frequent enough that a passive-vortex treatment becomes questionable. We use $\chi=0.1$ and $\chi=1$ as reference lines in the figures. They separate negligible, secular, and dynamically active regimes.

Each reconnection also transfers energy. The kinetic energy per unit length of a quantized vortex is
\begin{equation}
    \frac{E_\mathrm{line}}{\ell}\simeq \frac{\rho_\mathrm{c}\kappa^2}{4\pi}\ln\left(\frac{b}{\xi}\right),
    \label{eq:vortex_line_energy}
\end{equation}
where $b=\mathcal{L}^{-1/2}$ is the intervortex spacing and $\xi$ is the vortex-core or healing scale \citep{fet01,fet09}. Here $\rho_\mathrm{c}$ is the density participating in the coherent phase. It equals the core density in the zero-temperature SP benchmark; a finite-temperature SFDM application must replace it by the superfluid density $\rho_\mathrm{s}$.

In the non-interacting Schr\"odinger-Poisson limit, the vortex core lacks a fixed microscopic healing length. We therefore use Eq.~\eqref{eq:vortex_line_energy} as an extrapolation of the Gross-Pitaevskii line-energy estimate. Direct simulations of phase singularities are needed to test the logarithmic scaling in this limit.

One reconnection rearranges a line segment of order $b$. We define $\epsilon_\mathrm{rec}$ as the fraction of that segment's line energy transferred out of coherent vortex motion:
\begin{equation}
    \Delta E_\mathrm{rec}
    =\epsilon_\mathrm{rec}\frac{\rho_\mathrm{c}\kappa^2b}{4\pi}
    \ln\!\left(\frac{b}{\xi}\right).
    \label{eq:delta_e_rec}
\end{equation}
This definition keeps the logarithmic line-energy normalization explicit. Sound-emission calculations and reconnection measurements establish irreversible acoustic transfer in quantum fluids \citep{lea+01,sta+25}, but they do not calibrate its efficiency in a dark condensate. We use $\epsilon_\mathrm{rec}=10^{-2}$ as an illustrative fiducial value and leave its dependence explicit.

The corresponding volumetric heating estimate is
\begin{equation}
    Q_\mathrm{rec}\sim\frac{\Gamma_\mathrm{rec}V_\mathrm{core}\Delta E_\mathrm{rec}}{V_\mathrm{core}}
    =\Gamma_\mathrm{rec}\Delta E_\mathrm{rec}.
    \label{eq:qrec}
\end{equation}
The first equality shows the bookkeeping in the core volume. The second follows because $\Gamma_\mathrm{rec}$ is already a rate density. With $b=\mathcal{L}^{-1/2}$, the heating rate scales as $Q_\mathrm{rec}\propto\mathcal{L}^{2}$ at fixed local microphysics.

\subsection{Energetic impact on the core}

The energetic impact depends on the reservoir that receives or loses the reconnection energy. We compare $Q_\mathrm{rec}$ with two energy reservoirs. The first is the kinetic energy stored in the vortex tangle. Multiplying Eq.~\eqref{eq:vortex_line_energy} by the line density gives
\begin{equation}
    u_\mathrm{vort}\simeq
    \frac{\rho_\mathrm{c}\kappa^2\mathcal{L}}{4\pi}
    \ln\left(\frac{b}{\xi}\right).
    \label{eq:uvort}
\end{equation}
This is the standard logarithmic vortex-energy estimate used for dilute condensates and superfluids \citep{fet01,fet09,bar+23}. The time needed for reconnections to drain this reservoir is
\begin{equation}
    t_\mathrm{diss}\equiv \frac{u_\mathrm{vort}}{Q_\mathrm{rec}}
    \simeq
    \frac{1}{C_\mathrm{rec}\epsilon_\mathrm{rec}\kappa\mathcal{L}},
    \label{eq:tdiss}
\end{equation}
where we used $b=\mathcal{L}^{-1/2}$. The density and logarithmic line-energy factor cancel: $u_\mathrm{vort}$ and $\Delta E_\mathrm{rec}$ contain the same factors of $\rho_\mathrm{c}$ and $\ln(b/\xi)$. The cancellation does not mean that dense cores behave the same dynamically: density still enters $t_\mathrm{dyn}$, the soliton relation, and the possible thermal response of the dark medium. At fixed core size, lighter bosons have larger $\kappa$ and shorter $t_\mathrm{diss}$.

The second comparison asks whether the released energy can affect the bulk halo core. For a self-gravitating core the virial theorem gives an energy density of order \citep{bin08}
\begin{equation}
    u_\mathrm{vir}\sim \rho_\mathrm{c}\sigma_{\mathrm{c}}^2
    \sim G\rho_{\mathrm{c}}^2 r_\mathrm{soliton}^2,
    \label{eq:uvir}
\end{equation}
where $\sigma_c^2\sim G\rho_{\mathrm{c}} r_\mathrm{soliton}^2$ is the characteristic velocity dispersion. We define
\begin{equation}
    \Pi_\mathrm{rec}\equiv
    \frac{Q_\mathrm{rec}t_\mathrm{dyn}}{u_\mathrm{vir}}.
    \label{eq:pirec}
\end{equation}
Values $\Pi_\mathrm{rec}\ll1$ leave the bulk virial state nearly unchanged over one dynamical time. Values $\Pi_\mathrm{rec}\gtrsim1$ invalidate an energetically passive treatment of the assumed network, although they do not determine whether the core expands, loses mass, or exports the energy. A second dimensionless quantity measures the finite vortex reservoir,
\begin{equation}
    \beta_{\rm v}\equiv\frac{u_\mathrm{vort}}{u_\mathrm{vir}}
    =\Pi_\mathrm{rec}\frac{t_\mathrm{diss}}{t_\mathrm{dyn}}.
    \label{eq:beta_v}
\end{equation}
If no forcing regenerates vortex energy, the cumulative transfer over an interval $T$ obeys
\begin{equation}
    f_\mathrm{rec}(T)\equiv\frac{E_\mathrm{rec}(T)}{u_\mathrm{vir}V_\mathrm{core}}
    \leq\min\!\left[\Pi_\mathrm{rec}\frac{T}{t_\mathrm{dyn}},\,\beta_{\rm v}\right].
    \label{eq:cumulative_bound}
\end{equation}
The first term assumes that the initial rate persists; the second enforces the finite energy stored in an unforced tangle. Sustained forcing removes the second cap by replenishing the network, but then the forcing supplies the ultimate energy source and reconnection acts as the transfer channel.

\subsection{Angular-momentum normalization}

The angular momentum of a rotating vortex array provides an independent line-density estimate. A straight isolated vortex whose azimuthal flow is integrated out to a transverse cutoff radius $R_\perp$ carries
\begin{equation}
    \frac{L_z}{\ell}\simeq \frac{1}{2}\rho_{\mathrm{c}}\kappa R_\perp^2,
\end{equation}
up to the negligible core-radius correction. In a vortex lattice $R_\perp$ is naturally interpreted as the Wigner-Seitz-cell radius, while the net angular momentum of the rotating core is more cleanly encoded by the Feynman relation \citep{fey55} for a superfluid array with angular speed $\Omega$,
\begin{equation}
    n_v=\frac{2\Omega}{\kappa}.
\end{equation}
We use the core dynamical time $t_\mathrm{dyn}\sim(G\rho_{\mathrm{c}})^{-1/2}$, with $\rho_{\mathrm{c}}$ the core density. The dimensionless angular speed is $\eta_\Omega\equiv\Omega t_\mathrm{dyn}$. For any assumed line density, we write its normalization relative to the Mocz-scale estimate as $\eta_{\rm v}\equiv\mathcal{L}/\mathcal{L}_\mathrm{M}$. Treating the Feynman areal density $n_v$ as the line-density scale gives
\begin{equation}
    \mathcal{L}_\mathrm{F}\simeq \frac{2\eta_\Omega}{\kappa t_\mathrm{dyn}},
    \qquad
    \eta_{\rm v}^\mathrm{F}\equiv\frac{\mathcal{L}_\mathrm{F}}{\mathcal{L}_\mathrm{M}}
    =\frac{2\eta_\Omega(7.5r_{\mathrm{c}})^2}{\kappa t_\mathrm{dyn}}.
    \label{eq:feynman_line_density}
\end{equation}
The angular-speed parameter $\eta_\Omega$ maps to a line-density normalization $\eta_{\rm v}^\mathrm{F}$ once a rotation model and a core model are specified. For the non-interacting soliton relation used below, Eq.~\eqref{eq:feynman_line_density} gives $\eta_{\rm v}^\mathrm{F}\simeq8.4\eta_\Omega$. A slowly rotating core with $\eta_\Omega=0.03$--$0.1$ therefore has $\mathcal{L}_\mathrm{F}\simeq0.25$--$0.84\mathcal{L}_\mathrm{M}$. The fiducial Mocz-based normalization is consequently within an order of magnitude of a simple angular-momentum estimate, but the reconnection rate remains sensitive to this choice.

\subsection{Energy-release channels}

We use ``heating'' to mean energy transferred out of coherent vortex motion; it need not imply local thermodynamic equilibrium. Gross-Pitaevskii calculations produce compressible waves near the reconnection \citep{lea+01,pro20}, Kelvin waves \citep{min+25,sco+26}, and vortex loops \citep{zuc+12,vil+17,and+18}. In a dark condensate, the corresponding products can include dark phonons \citep{bal+22}, density waves, Kelvin-wave cascades, loops, and escaping scalar waves analogous to gravitational cooling \citep{guz06}.

Minimal BECDM carries no electromagnetic or weak charge, so all of this energy remains in the dark sector. Photons or neutrinos require additional portal physics such as a dark $U(1)$, kinetic mixing, a millicharge, a baryon coupling, or a neutrino portal \citep{ack+09,fan+13,ros17,cap+20}. Phonon-mediated SFDM already couples the condensate to baryons \citep{ber15,ber16,ber+18,mis19}, but the coupling does not turn $Q_\mathrm{rec}$ directly into an ordinary luminosity. A visible signal requires a separate transport and cooling calculation. Individual nonrelativistic reconnections also radiate gravitational waves with negligible efficiency \citep{mag07}. We therefore set Standard-Model luminosities to zero and use the dark-sector energy budget as the conservative phenomenological constraint.

\section{Results}
\label{sec:results}

We scan the parameter range
\begin{equation}
10^{-24}\,\mathrm{eV}\leq m\leq10^{-20}\,\mathrm{eV},
\qquad
10^6\leq\rho_{\mathrm{c}}/(M_\odot\,\mathrm{kpc}^{-3})\leq10^9,
\end{equation}
and use $C_\mathrm{rec}=0.1,1,10$ to show the uncertainty in the reconnection-rate normalization. Unless stated otherwise, the displayed heat map uses $C_\mathrm{rec}=1$, $r_{\mathrm{c}}=1\,\mathrm{kpc}$, $\epsilon_\mathrm{rec}=10^{-2}$, and $\eta_{\rm v}=1$. This fixed-radius scan is a controlled stress test for externally specified cores, not a prediction for every relaxed fuzzy-dark-matter halo. Equilibrium non-interacting Schr\"odinger-Poisson solitons obey a separate relation among $m$, $\rho_{\mathrm{c}}$, and $r_{\mathrm{c}}$.

From this scan we record $t_\mathrm{rec}$, the event-count parameter $\chi=t_\mathrm{dyn}/t_\mathrm{rec}$, the dark-sector heating rate $Q_\mathrm{rec}$, and the energy diagnostics in Eqs.~\eqref{eq:tdiss}, \eqref{eq:pirec}, and \eqref{eq:beta_v}. These quantities separate frequent reconnections from energetically important reconnections. For the fiducial non-interacting branch we set $\ln(b/\xi)=3$, consistent with the estimate $b/\xi\sim10$--$15$ discussed below. Changing this factor rescales $u_\mathrm{vort}$, $Q_\mathrm{rec}$, $\Pi_\mathrm{rec}$, and $\beta_{\rm v}$ together, while $t_\mathrm{diss}$ remains unchanged. The non-interacting Schr\"odinger-Poisson branch may instead have an effective cutoff tied to the de Broglie scale, so this logarithm should not be read as a measured microscopic quantity.

At fixed $r_{\mathrm{c}}$ the scaling is transparent. Since $\mathcal{L}\propto d_\mathrm{peak}^{-2}\propto r_{\mathrm{c}}^{-2}$ and $V_\mathrm{core}\propto r_{\mathrm{c}}^3$, the total rate scales as $\Gamma_\mathrm{rec}V_\mathrm{core}\propto C_\mathrm{rec}\kappa r_{\mathrm{c}}^{-2}$. Thus
\begin{equation}
    \chi\propto C_\mathrm{rec}m^{-1}r_{\mathrm{c}}^{-2}\rho_{\mathrm{c}}^{-1/2}.
\end{equation}
The energetic diagnostics scale as
\begin{equation}
    \frac{t_\mathrm{diss}}{t_\mathrm{dyn}}
    \propto
    \frac{m r_{\mathrm{c}}^2\rho_{\mathrm{c}}^{1/2}}{C_\mathrm{rec}\epsilon_\mathrm{rec}},
    \qquad
    \Pi_\mathrm{rec}
    \propto
    C_\mathrm{rec}\epsilon_\mathrm{rec}\ln(b/\xi)
    m^{-3}r_{\mathrm{c}}^{-6}\rho_{\mathrm{c}}^{-3/2}.
    \label{eq:diagnostic_scalings}
\end{equation}
Lighter particles have larger circulation and hence more frequent reconnections in the fixed-radius survey. They also have shorter vortex-dissipation times and a larger possible virial impact. More compact cores increase all reconnection effects. Increasing the core density shortens the dynamical time; at fixed $r_{\mathrm{c}}$ it decreases both $\chi$ and $\Pi_\mathrm{rec}$.

\begin{figure}[!htbp]
\centering
\includegraphics[width=0.95\columnwidth]{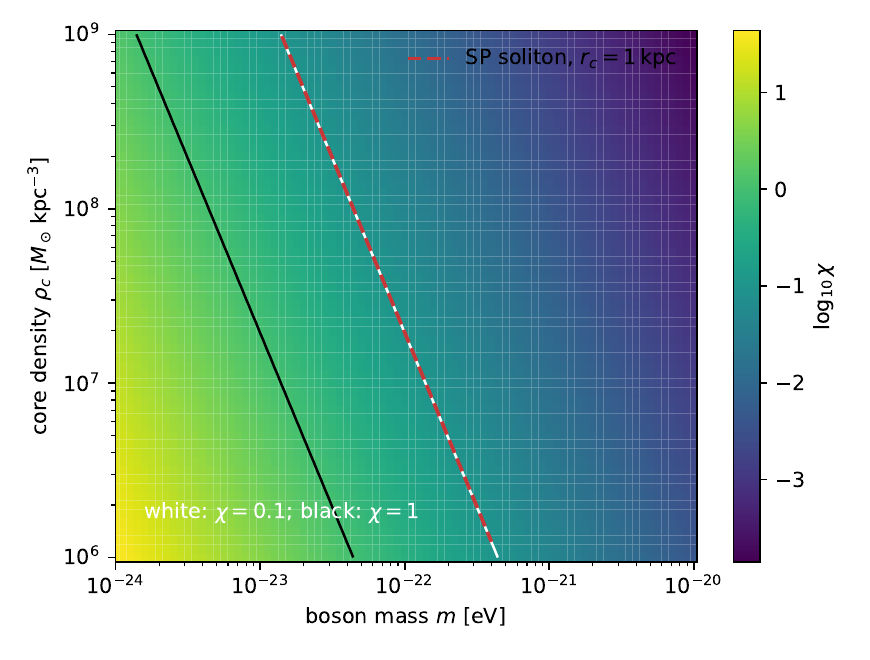}
\caption{Fixed-radius semi-analytical scan for $C_\mathrm{rec}=1$, $r_{\mathrm{c}}=1\,\mathrm{kpc}$, and the Mocz normalization $\eta_{\rm v}=1$. The color shows $\log_{10}\chi$, where $\chi=t_\mathrm{dyn}/t_\mathrm{rec}$. The white and black contours mark $\chi=0.1$ and $\chi=1$. The dashed red curve shows the non-interacting soliton relation of Eq.~\eqref{eq:schive_relation} at $r_{\mathrm{c}}=1\,\mathrm{kpc}$; it tracks the white contour because the soliton relation gives $\chi\simeq 0.10$ for $\eta_{\rm v}=1$. For smaller line densities, $\Delta\log_{10}\chi=(5/2)\log_{10}\eta_{\rm v}<0$, so the contours move toward smaller $\chi$; the values in Table~\ref{tab:diagnostic_summary} should be rescaled in the same way.}
\label{fig:heating_regimes}
\end{figure}

\autoref{fig:heating_regimes} shows $\log_{10}\chi$ in the $(m,\rho_{\mathrm{c}})$ plane for $C_\mathrm{rec}=1$ and $\eta_{\rm v}=1$. The white curve marks $\chi=0.1$ and the black curve marks $\chi=1$. The dashed red curve marks the non-interacting soliton relation evaluated at $r_{\mathrm{c}}=1\,\mathrm{kpc}$. Its near-overlap with the white contour is expected: imposing the soliton relation cancels the explicit $m$ and $\rho_{\mathrm{c}}$ dependence of $\chi$ and gives the Mocz-normalized value $\chi\simeq0.10$. For the fixed-radius scan at $\eta_{\rm v}=1$ we find $1.4\times10^{-4}\lesssim\chi\lesssim 4.4\times10^{1}$. The threshold $\chi=1$ occurs near $m\simeq4\times10^{-23}\,\mathrm{eV}$ at $\rho_{\mathrm{c}}=10^6M_\odot\,\mathrm{kpc}^{-3}$, but near $m\simeq 4\times10^{-24}\,\mathrm{eV}$ at $\rho_{\mathrm{c}}=10^8M_\odot\,\mathrm{kpc}^{-3}$. The displayed contours shift as $C_\mathrm{rec}\eta_{\rm v}^{5/2}$.

The fixed-radius energetic diagnostics sharpen the interpretation. Across the same scan, and for $\ln(b/\xi)=3$, we obtain
\begin{equation}
    0.97\lesssim \frac{t_\mathrm{diss}}{t_\mathrm{dyn}}
    \lesssim 3.1\times10^5,
    \qquad
    3.8\times10^{-13}\lesssim \Pi_\mathrm{rec}
    \lesssim 1.2\times10^4.
\end{equation}
The large upper end of $\Pi_\mathrm{rec}$ occurs at the lightest masses and lowest core densities in the fixed-radius scan. In that corner, the inferred vortex energy can exceed the simple virial estimate, so the calculation flags a failure of the passive-vortex bookkeeping for the assumed $\eta_{\rm v}=1$ core. It does not predict core destruction. At $\rho_{\mathrm{c}}=10^8M_\odot\,\mathrm{kpc}^{-3}$, for example, $m=10^{-24}\,\mathrm{eV}$ gives $t_\mathrm{diss}/t_\mathrm{dyn}\simeq9.7$ and $\Pi_\mathrm{rec}\simeq12$, while $m=4\times10^{-23}\,\mathrm{eV}$ gives $t_\mathrm{diss}/t_\mathrm{dyn}\simeq3.9\times10^2$ and $\Pi_\mathrm{rec}\simeq1.9\times10^{-4}$. Thus the reconnection count can become dynamical over a wider region than the region where the released energy competes with the virial reservoir.

For a relaxed soliton, the core relation changes this interpretation. We use the standard relation \citep{sch+14}
\begin{equation}
    \rho_{\mathrm{c}}\simeq 1.95\times10^7
    \left(\frac{m}{10^{-22}\,\mathrm{eV}}\right)^{-2}
    \left(\frac{r_{\mathrm{c}}}{1\,\mathrm{kpc}}\right)^{-4}
    M_\odot\,\mathrm{kpc}^{-3}.
    \label{eq:schive_relation}
\end{equation}
The associated fitted density profile is
\begin{equation}
    \rho_\mathrm{sol}(r)=\rho_\mathrm{c}
    \left[1+0.091\left(\frac{r}{r_\mathrm{c}}\right)^2\right]^{-8}.
    \label{eq:schive_profile}
\end{equation}
Only two of $m$, $\rho_{\mathrm{c}}$, and $r_{\mathrm{c}}$ are then independent. Equivalently,
\begin{equation}
    r_{\mathrm{c}}\simeq
    \left[\frac{1.95\times10^7}{\rho_{\mathrm{c}}/(M_\odot\,\mathrm{kpc}^{-3})}
    \left(\frac{m}{10^{-22}\,\mathrm{eV}}\right)^{-2}\right]^{1/4}\mathrm{kpc}.
\end{equation}
Substituting this relation into Eq.~\eqref{eq:diagnostic_scalings} cancels the explicit $m$ and $\rho_{\mathrm{c}}$ dependence:
\begin{equation}
    \chi\propto C_\mathrm{rec},
    \qquad
    \frac{t_\mathrm{diss}}{t_\mathrm{dyn}}\propto
    \frac{1}{C_\mathrm{rec}\epsilon_\mathrm{rec}},
    \qquad
    \Pi_\mathrm{rec}\propto C_\mathrm{rec}\epsilon_\mathrm{rec}\ln(b/\xi).
\end{equation}
With $C_\mathrm{rec}=1$, $\epsilon_\mathrm{rec}=10^{-2}$, $\ln(b/\xi)=3$, $r_\mathrm{soliton}=3.5r_{\mathrm{c}}$, and $d_\mathrm{peak}=7.5r_{\mathrm{c}}$, the constants are
\begin{equation}
    \chi\simeq0.10,
    \qquad
    \frac{t_\mathrm{diss}}{t_\mathrm{dyn}}\simeq4.3\times10^2,
    \qquad
    \Pi_\mathrm{rec}\simeq1.4\times10^{-4},
    \qquad
    \beta_{\rm v}\simeq6.0\times10^{-2}.
\end{equation}
The main numerical outcomes for $\eta_{\rm v}=1$ are summarized in \autoref{tab:diagnostic_summary}.

\begin{table}[!htbp]
\centering
\small
\setlength{\tabcolsep}{5pt}
\renewcommand{\arraystretch}{1.12}
\begin{tabular}{p{0.50\textwidth}ll}
\toprule
Core model & Diagnostic & Value \\
\midrule
Fixed $r_{\mathrm{c}}=1\,\mathrm{kpc}$ scan
& $\chi$ & $1.4\times10^{-4}$--$4.4\times10^{1}$ \\
& $t_\mathrm{diss}/t_\mathrm{dyn}$ & $0.97$--$3.1\times10^{5}$ \\
& $\Pi_\mathrm{rec}$ & $3.8\times10^{-13}$--$1.2\times10^{4}$ \\
& $\beta_{\rm v}$ & $1.2\times10^{-7}$--$1.2\times10^{4}$ \\
\addlinespace
SP soliton with $\eta_{\rm v}=1$
& $\chi$ & $0.10$ \\
& $t_\mathrm{diss}/t_\mathrm{dyn}$ & $4.3\times10^{2}$ \\
& $\Pi_\mathrm{rec}$ & $1.4\times10^{-4}$ \\
& $\beta_{\rm v}$ & $6.0\times10^{-2}$ \\
\addlinespace
SP soliton with $\eta_\Omega=0.03$--$0.1$
& $\chi$ & $3.1\times10^{-3}$--$6.4\times10^{-2}$ \\
& $t_\mathrm{diss}/t_\mathrm{dyn}$ & $5.1\times10^{2}$--$1.7\times10^{3}$ \\
& $\Pi_\mathrm{rec}$ & $8.8\times10^{-6}$--$9.9\times10^{-5}$ \\
& $\beta_{\rm v}$ & $1.5\times10^{-2}$--$5.0\times10^{-2}$ \\
\bottomrule
\end{tabular}
\caption{Conditional diagnostics for the fixed-radius scan and the non-interacting Schr\"odinger-Poisson (SP) soliton. All rows use $C_\mathrm{rec}=1$, $\epsilon_\mathrm{rec}=10^{-2}$, and $\ln(b/\xi)=3$. The fixed-radius rows report the full range at $\eta_{\rm v}=1$; values with $\beta_{\rm v}\gtrsim1$ are self-inconsistent rather than viable predictions. The last group uses $\eta_{\rm v}=\eta_{\rm v}^{\rm F}\simeq0.25$--$0.84$ from Eq.~\eqref{eq:feynman_line_density}. A vortex-free core has zero reconnection rate.}
\label{tab:diagnostic_summary}
\end{table}

For the fiducial microphysics, the formal SP continuum extrapolation gives one reconnection per ten dynamical times and transfers $1.4\times10^{-4}$ of the virial energy per dynamical time. The heating time $u_\mathrm{vir}/Q_\mathrm{rec}=t_\mathrm{dyn}/\Pi_\mathrm{rec}$ is $7.1\times10^3t_\mathrm{dyn}$. The vortex reservoir depletes after $4.3\times10^2t_\mathrm{dyn}$ and stores $\beta_{\rm v}\simeq0.060$ of the virial energy. These continuum values describe a hypothetical disturbed ensemble; Eq.~\eqref{eq:vortex_crossings} shows that they do not describe a confirmed tangle in one relaxed core.

For the benchmark $m=10^{-22}\,\mathrm{eV}$ and $r_\mathrm{c}=1\,\mathrm{kpc}$, the soliton relation gives $t_\mathrm{dyn}=1.07\times10^8\,\mathrm{yr}$. Equation~\eqref{eq:cumulative_bound} then limits the energy transferred by an unforced $\eta_{\rm v}=1$ tangle over $10\,\mathrm{Gyr}$ to
\begin{equation}
    f_\mathrm{rec}(10\,\mathrm{Gyr})\leq1.3\times10^{-2}.
    \label{eq:ten_gyr_bound}
\end{equation}
The Feynman range $\eta_{\rm v}^{\rm F}=0.25$--$0.84$ lowers this bound to $8.2\times10^{-4}$--$9.3\times10^{-3}$. A tangle maintained at a constant rate for $10\,\mathrm{Gyr}$ reaches an order-unity virial transfer only for $\eta_{\rm v}\gtrsim8.7$; an unforced reservoir contains that much energy only for $\eta_{\rm v}\gtrsim17$. Both thresholds lie above the Mocz normalization and require several vortex crossings of the core. The published spectral peak \citep{moc+17} alone cannot establish either line density.

\begin{figure}[!htbp]
\centering
\includegraphics[width=0.95\columnwidth]{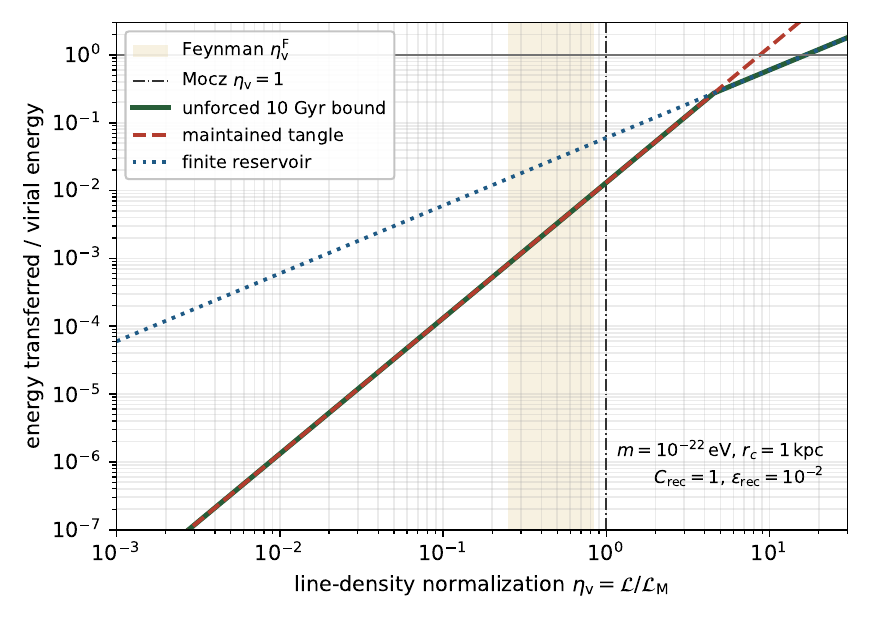}
\caption{Cumulative reconnection-energy bound for an SP soliton with $m=10^{-22}\,\mathrm{eV}$, $r_\mathrm{c}=1\,\mathrm{kpc}$, $C_\mathrm{rec}=1$, $\epsilon_\mathrm{rec}=10^{-2}$, and $\ln(b/\xi)=3$. The curves are formal continuum extrapolations; at $\eta_{\rm v}=1$ the expected number of projected crossings is below unity. The dashed curve extrapolates a tangle maintained at its initial line density for $10\,\mathrm{Gyr}$. The dotted curve is the total energy initially stored in the vortex reservoir. The solid curve is the unforced bound, the smaller of these two quantities. The shaded band gives the Feynman range $\eta_{\rm v}^{\rm F}=0.25$--$0.84$, and the vertical dash-dotted line marks the Mocz normalization $\eta_{\rm v}=1$. The horizontal line denotes energy transfer equal to the virial scale.}
\label{fig:line_density_sensitivity}
\end{figure}

Along the SP soliton sequence, the full fiducial rescaling is
\begin{align}
    \chi &\simeq0.10\,C_\mathrm{rec}\eta_{\rm v}^{5/2},\\
    \frac{t_\mathrm{diss}}{t_\mathrm{dyn}}
    &\simeq4.3\times10^2 C_\mathrm{rec}^{-1}
    \left(\frac{\epsilon_\mathrm{rec}}{10^{-2}}\right)^{-1}\eta_{\rm v}^{-1},\\
    \Pi_\mathrm{rec}
    &\simeq1.4\times10^{-4} C_\mathrm{rec}
    \left(\frac{\epsilon_\mathrm{rec}}{10^{-2}}\right)
    \left(\frac{\ln(b/\xi)}{3}\right)\eta_{\rm v}^{2},\\
    \beta_{\rm v}
    &\simeq6.0\times10^{-2}
    \left(\frac{\ln(b/\xi)}{3}\right)\eta_{\rm v}.
    \label{eq:soliton_rescaling}
\end{align}
Figure~\ref{fig:line_density_sensitivity} shows the resulting ten-Gyr energy bound. Below the reservoir-limited turnover, the transferred fraction scales as $\eta_{\rm v}^2$; after depletion it cannot exceed the finite reservoir, which scales as $\eta_{\rm v}$.

\subsection{Aperture-matched dwarf sample}
\label{sec:dwarf_population}

We next translate the bound to measured systems. Wolf et al. \citep{wol+10} give deprojected half-light radii $r_{1/2}$ and dynamical masses $M_{1/2}$ for the classical Milky Way dwarf spheroidals using one mass estimator. We use their values for Fornax, Sculptor, Carina, Draco, Leo~I, Leo~II, Sextans, and Ursa Minor. At fixed $m=10^{-22}\,\mathrm{eV}$, we determine one conditional SP core radius for each system from
\begin{equation}
    4\pi\int_0^{r_{1/2}}\rho_\mathrm{sol}(r)r^2\,dr=M_{1/2},
    \label{eq:aperture_match}
\end{equation}
with Eq.~\eqref{eq:schive_profile}. This construction uses the full measured dynamical mass, without subtracting the stellar contribution, and assumes that the soliton supplies all of it. It is an aperture-mass translation of the reconnection bound, not a new fit or evidence that these galaxies contain SP solitons. Keeping one homogeneous dispersion-supported sample also avoids mixing half-light masses with model-dependent rotation-curve core fits. The Wolf et al.\ estimator itself carries an intrinsic $1\sigma$ systematic scatter of order $20$--$25\%$, dominated by the unknown three-dimensional shape of the stellar distribution rather than by the quoted observational errors on $R_e$ and $\langle\sigma_\mathrm{los}\rangle$ \citep{wol+10,cam+17}; propagated through $\rho_\mathrm{c}\propto r_\mathrm{c}^{-4}$ and $t_\mathrm{dyn}\propto\rho_\mathrm{c}^{-1/2}$ at fixed $m$, this corresponds to a comparable ($\sim 40$--$50\%$) systematic uncertainty in $t_\mathrm{dyn}$, and hence in $f_\mathrm{rec}(10\,\mathrm{Gyr})$, without altering the conclusion that the population ceiling remains far below the virial-energy scale.
We note that the population ceiling is insensitive to the fiducial choice of $m$ in this sense: $\beta_v$, which caps the bound whenever the tangle is reservoir-limited, does not depend on $m$ or $r_\mathrm{c}$ along the soliton sequence.

\begin{figure}[!htbp]
\centering
\includegraphics[width=0.98\textwidth]{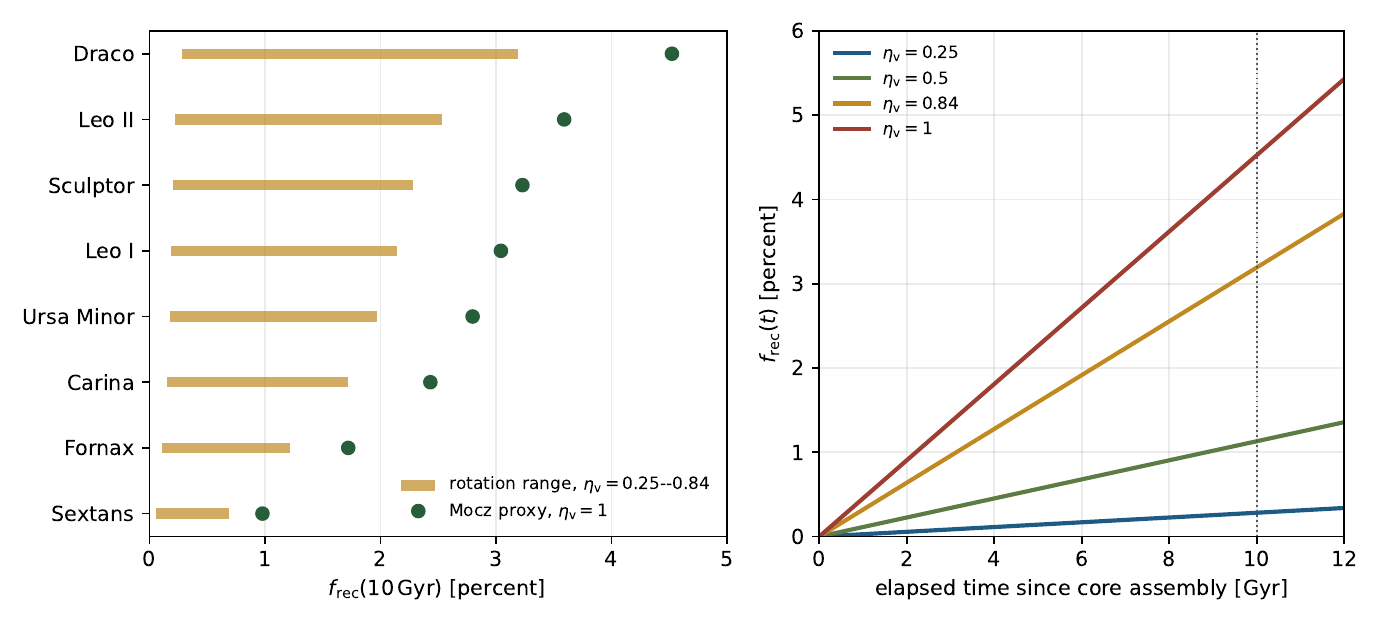}
\caption{Conditional population mapping and time dependence for $m=10^{-22}\,\mathrm{eV}$, $C_\mathrm{rec}=1$, $\epsilon_\mathrm{rec}=10^{-2}$, and $\ln(b/\xi)=3$. (a) Each point uses the Wolf et al. \citep{wol+10} half-light radius and enclosed dynamical mass, with an SP soliton assigned through Eq.~\eqref{eq:aperture_match}. Green points use the Mocz proxy $\eta_{\rm v}=1$; gold bars span the modest-rotation normalization $\eta_{\rm v}=0.25$--$0.84$. The bars show model normalization, not observational error bars. (b) Cumulative unforced bound for the Draco aperture-matched soliton, which has the shortest dynamical time in this sample. The dotted vertical line marks $10\,\mathrm{Gyr}$. Both panels remain conditional on the existence of a disordered vortex population.}
\label{fig:dwarf_population}
\end{figure}

The inferred radii span $r_\mathrm{c}=0.54$--$1.16\,\mathrm{kpc}$ and the central densities span $1.08\times10^7$--$2.32\times10^8M_\odot\,\mathrm{kpc}^{-3}$. \autoref{fig:dwarf_population}(a) shows the resulting ten-Gyr bounds. At the Mocz normalization they range from $0.98\%$ for Sextans to $4.52\%$ for Draco. The rotation-based interval $\eta_{\rm v}=0.25$--$0.84$ gives $0.061$--$3.19\%$ across the sample. The associated fixed-structure velocity ceilings remain below $2.24\%$ and $1.58\%$, respectively. Panel (b) follows the most compact mapped system, Draco, from core assembly to $12\,\mathrm{Gyr}$. The transfer remains gradual throughout the plotted interval; at $10\,\mathrm{Gyr}$ it is $0.28\%$, $3.19\%$, and $4.52\%$ for $\eta_{\rm v}=0.25$, $0.84$, and $1$.

Dwarf rotation-curve diversity motivates tests of processes that can alter inner halo structure \citep{oma+15}. In collisionless-CDM models, repeated potential fluctuations can transfer energy to dark-matter orbits and turn cusps into cores \citep{pon12}; estimates for dwarf-spheroidal halos require roughly $10^{53}$--$10^{55}\,\mathrm{erg}$, depending on halo mass and the desired core size \citep{pen+12}. An absolute-energy overlay with the present bound would not be a like-for-like comparison. Those studies inject energy into an initially cusped CDM halo, whereas reconnection redistributes energy stored in vortices inside an already-cored SP solution, and our virial normalization is defined for that solution. Within the present framework the controlled comparison is therefore dimensionless: the largest population ceiling is $0.0452E_\mathrm{vir}$, at least a factor of 22 below the virial-energy scale associated with an order-unity rearrangement. Reconnection cannot by itself act as a competitive core-formation mechanism under the fiducial assumptions.

\section{Phenomenological consequences and limitations}
\label{sec:discussion}

\paragraph{Kinematic bound:} The aperture-matched sample turns Eq.~\eqref{eq:cumulative_bound} into a population-level ceiling. If all transferred energy entered the same kinetic reservoir while the mass and radius remained fixed, $u_\mathrm{vir}\propto\sigma_c^2$ would give
\begin{equation}
    \frac{|\Delta\sigma_c|}{\sigma_c}
    \leq\sqrt{1+f_\mathrm{rec}}-1.
    \label{eq:velocity_bound}
\end{equation}
The largest formal change is $2.24\%$ at $\eta_{\rm v}=1$ and $1.58\%$ over the rotation-based range. Core expansion or energy escape would reduce the local change, so this is a fixed-structure ceiling rather than a predicted dispersion or rotation curve. A larger kinematic evolution requires another process or a denser, continuously forced tangle. In the latter case the driver supplies the energy and reconnection is only the transfer channel.

\paragraph{Core model and vortex identity:} Equation~\eqref{eq:schive_relation} applies to a non-interacting SP soliton. The fixed-radius map in Fig.~\ref{fig:heating_regimes} is a stress test for perturbed, externally confined, or self-interacting cores, not an equilibrium fuzzy-dark-matter sequence. A branch-specific mass-radius relation must replace Eq.~\eqref{eq:schive_relation} in SFDM. The vortex line energy also assumes a compact Gross-Pitaevskii core with outer cutoff $b$. A virial estimate gives
\begin{equation}
    \frac{\lambda_\mathrm{dB}}{d_\mathrm{peak}}
    \sim\frac{\kappa}{7.5r_\mathrm{c}r_\mathrm{soliton}(G\rho_\mathrm{c})^{1/2}},
\end{equation}
where $\lambda_\mathrm{dB}=h/(m\sigma_c)$ and $\sigma_c^2\sim G\rho_c r_\mathrm{soliton}^2$. Along the SP soliton sequence this ratio is about $0.5$. Taking $\xi\sim\lambda_\mathrm{dB}/2\pi$ gives $b/\xi\sim10$--$15$ and motivates the fiducial $\ln(b/\xi)=3$. Identifying $\xi$ with the full de Broglie scale would erase the separation needed for the logarithmic line-tension formula. The SP result is therefore an order-of-magnitude bound on nodal-line energy, while the formula is better controlled in a contact-interacting condensate with a distinct healing length.

\paragraph{Secondary observables:} Any structural response, stellar heating, or lensing perturbation driven by reconnection must fit inside the population bounds in Fig.~\ref{fig:dwarf_population}. Existing stellar-heating calculations follow soliton oscillations and stochastic granules rather than vortex dissipation \citep{li+21,chi+21}; applying them here requires the spatial and temporal spectrum of the reconnection-driven gravitational potential, not only its total energy. The same missing transfer function prevents a lensing prediction. A haloscope coherence time is also a different quantity: it is set by the local field's velocity distribution and spectral linewidth, whereas $t_\mathrm{rec}$ counts events integrated over the whole model core. Treating one core-wide event interval as a detector coherence time would mix two observables and, for an ADMX-style comparison, two very different particle-mass ranges. The present calculation therefore constrains core energetics and longevity but does not predict a haloscope linewidth.

\paragraph{Phonon-mediated SFDM:} A branch-specific thermal criterion can be written for SFDM, but the SP normalization cannot supply its inputs. For one linear phonon mode with sound speed $c_s$,
\begin{equation}
    u_\phi(T)=\int_0^T c_V(T')\,dT'
    =\frac{\pi^2k_B^4T^4}{30\hbar^3c_s^3}.
\end{equation}
If a fraction $f_\phi$ of $Q_\mathrm{rec}$ thermalizes in that mode, the illustrative SP-normalized timescale is
\begin{equation}
    \frac{t_T}{t_\mathrm{dyn}}
    \simeq5.6\times10^{-2}f_\phi^{-1}
    \left(\frac{T_c}{1\,\mathrm{mK}}\right)^4
    \left(\frac{c_s}{100\,\mathrm{km\,s^{-1}}}\right)^{-3}
    \left(\frac{Q_\mathrm{rec}}{5.6\times10^{-32}\,\mathrm{W\,m^{-3}}}\right)^{-1}
    \left(\frac{t_\mathrm{dyn}}{1.1\times10^8\,\mathrm{yr}}\right)^{-1}.
    \label{eq:thermal_feedback_time}
\end{equation}
This equation defines a test, not an SFDM prediction. The SFDM equation of state must provide the superfluid density, $T_c$, $c_s$, the phonon dispersion, and $f_\phi$, while its vortex population fixes $Q_\mathrm{rec}$. If $\rho_\mathrm{s}<\rho_\mathrm{c}$, both $Q_\mathrm{rec}$ and $u_\mathrm{vort}$ acquire the factor $\rho_\mathrm{s}/\rho_\mathrm{c}$; their ratio $t_\mathrm{diss}$ is unchanged. The present non-interacting calculation therefore favors neither BECDM nor SFDM as a dark-matter theory. It constrains the reconnection channel in the SP branch and gives the quantities that a separate SFDM calculation must replace.

\section{Conclusions}
\label{sec:conclusions}

We have derived a conditional upper bound on energy transfer by vortex reconnections in condensed dark-matter cores. The calculation combines a local quantum-fluid reconnection law with an explicit line-density normalization and caps the cumulative transfer by the finite energy initially stored in an unforced network.

For the benchmark $m=10^{-22}\,\mathrm{eV}$ and $r_\mathrm{c}=1\,\mathrm{kpc}$ SP soliton, the ten-Gyr ceiling is $1.3\%$ of the virial energy. Matching the same soliton profile to the homogeneous Wolf et al. half-light masses for eight classical dwarf spheroidals gives $0.98$--$4.52\%$ at the Mocz line-density proxy and $0.061$--$3.19\%$ for the modest-rotation normalization. The corresponding fixed-structure velocity changes stay below $2.24\%$ and $1.58\%$. Even the largest population value is at least a factor of 22 below the core virial-energy scale, so reconnection is not a competitive core-formation mechanism under the fiducial assumptions.

The population mapping is deliberately conditional. It assigns each measured aperture mass to an SP soliton at fixed particle mass, and the line density still comes from a turbulent spectral proxy rather than a phase-singularity count. A relaxed vortex-free soliton has no reconnection heating. Conversely, order-unity transfer over $10\,\mathrm{Gyr}$ requires a maintained tangle with $\eta_{\rm v}\gtrsim8.7$, well above the reference normalization; then the mechanism that maintains the tangle is also part of the energy budget.

The result is a constraint on core energetics and longevity, not a photon luminosity, a lensing map, or a haloscope linewidth. Those observables require branch-specific response and coupling models. The numerical population result applies to the non-interacting SP branch. A corresponding SFDM calculation must provide its own core relation, vortex-line density, superfluid fraction, and phonon thermalization model. Measuring the line density directly is therefore the decisive next test of whether reconnections are merely present or dynamically relevant.

\acknowledgments

The author thanks A.~Sava{\c s} Arapo\u{g}lu for his careful reading of the earlier version of the manuscript and useful comments. The author has used generative artificial intelligence (GenAI) to check spelling, grammar, and punctuation in this document.


\bibliographystyle{JHEP}
\bibliography{sfdm}

\end{document}